\documentclass[11pt]{iopart}
\pdfoutput=1
\usepackage{epsfig,graphicx,latexsym,iopams}
\def\rmd{{\rm d}}

\def\MNRAS{{\em Mon. Not. Roy. Astron. Soc. }}
\def\CQG{{\em Class. Quantum Grav. }}
\def\PRD{{\em Phys. Rev. D }}
\def\apj{{\em Astrophys. J.}}
\newcommand{\erf}[1]{(\ref{#1})}

\begin{document}

\title{Probing black holes at low redshift using LISA EMRI observations}
\author{Jonathan R Gair}
\address{Institute of Astronomy, University of Cambridge, Madingley 
Road, Cambridge, CB3 0HA, UK}
\ead{jgair@ast.cam.ac.uk}

\date{\today}

\begin{abstract}
One of the most exciting potential sources of gravitational waves for the Laser Interferometer Space Antenna (LISA) are the inspirals of approximately solar mass compact objects into massive black holes in the centres of galaxies --- extreme mass ratio inspirals (EMRIs). LISA should observe between
a few tens and a few hundred EMRIs over the mission lifetime, mostly at low redshifts ($z \lesssim1$). Each observation will provide a measurement of the parameters of the host system to unprecendented precision. LISA EMRI observations will thus offer a new and unique way to probe black holes at low redshift. In this article we provide a description of the population of EMRI events that LISA is likely to observe, and describe how the numbers of events vary with changes in the underlying assumptions about the black hole population. We also provide fitting functions that characterise LISA's ability to detect EMRIs and which will allow LISA event rates to be computed for arbitrary population models. We finish with a discussion of an ongoing programme that will use these results to assess what constraints LISA observations could place on galaxy evolution models.
\end{abstract}

\maketitle

\section{Introduction}
\label{intro}
The inspiral of a stellar mass compact object --- a black hole (BH), neutron star (NS) or white dwarf (WD) --- into a massive black hole (MBH) with mass in the range $10^4M_{\odot}$--$10^7M_{\odot}$ in the centre of a galaxy will generate gravitational waves at frequencies to which the planned space based gravitational wave detector, LISA~\cite{LISAppa} will be sensitive. These ``extreme mass ratio inspiral'' (EMRI) sources are of particular interest to relativists as the emitted gravitational waves encode detailed information about the spacetime structure close to the massive central object and hence can be used to test whether these objects are indeed the Kerr black holes predicted by relativity (see~\cite{Pau} and references therein). However, these sources are also of great interest for astrophysics. A LISA observation of an EMRI event can determine the parameters of the system to very high precision --- the mass of the inspiraling object and the mass and spin of the central black hole are typically recovered to accuracies of a fraction of a per cent~\cite{AK}. Preliminary estimates have indicated that LISA may see as many as a few thousand EMRI events over the mission lifetime~\cite{gairetal}. This set of observed EMRI events will provide us with detailed information about the properties of black holes in the relatively nearby Universe.

To date, EMRI event rate calculations have been done only crudely and for specific models of the black hole population. In addition, there is no simple prescription in the literature to go from an astrophysical population model to a realisation of the LISA event distribution. In this paper, we will describe some of the properties of the likely set of LISA EMRI events more carefully. In order to do so, we have computed the distance to which LISA will see different types of events and have folded these ranges into a simple population model to calculate the expected number and distribution of observed events. In the course of this work, we have derived fits which give the detectable depth of an EMRI event as a function of its parameters and we also provide these here. We present EMRI event distributions for a simple set of MBH population models in which there is no evolution over the redshift range to which LISA is sensitive ($0 < z < 1.5$). These models illustrate the potential for astrophysics that LISA EMRI observations offer, while the fits will enable other researchers to examine alternative galactic black hole population models and readily convert these into LISA event distributions. 

The paper is organised as follows. In Section~\ref{LISAsens} we describe the waveform model we have used to compute LISA signal-to-noise ratios and provide fits to the ``observable lifetime'' of events, which characterises the likelihood that LISA could see them. In Section~\ref{LISAevents} we use these results to compute event rates assuming a simple non-evolving MBH population. We compute the distribution of LISA events, and describe how the number of events and the minimum and maximum likely redshift of observed events vary with the inspiral rate and the spin of the MBH. Finally, in Section~\ref{app} we discuss how we are currently using these results to investigate LISA's ability to distinguish between different galaxy evolution models, as well as describing some of the limitations of the current work and how we plan to address these in the future.

\section{LISA sensitivity to EMRIs}
\label{LISAsens}
To compute the population of LISA EMRI events, we need a prescription to determine when an event is detectable. Typically, we expect an event to be detected if the matched filtering signal-to-noise ratio (SNR) exceeds some predetermined threshold, $\rho_{\rm thresh}$. The first estimate of the SNR threshold that will be needed was $\rho_{\rm thresh} \approx 35$, which was computed assuming that the data analysis would be carried out using a hierarchical semi-coherent matched filtering algorithm~\cite{gairetal}. This is somewhat higher than the threshold estimated to be needed for a fully coherent search, $\rho_{\rm thresh} \approx 15$, if such a search was computationally feasible. Several alternative approaches to EMRI data analysis have been proposed subsequently, including time-frequency analyses~\cite{wengair05,gairwen05,gairjones06,EtfAG} and Metropolis-Hastings Monte Carlo (MHMC) searches~\cite{stroeer06,neilemri,BBGP}. Time-frequency analyses will require higher SNRs for detections. MHMC searches may be able to get closer to the fully-coherent limit of $\rho_{\rm thresh} \approx 15$, but this has not yet been demonstrated for isolated sources let alone multiple overlapping sources. In this work we will take $\rho_{\rm thresh} = 30$ and assume that the LISA mission lasts $5$ years, with no failures on the satellite so that the two-independent low-frequency channels~\cite{cutler98} are available for the entire mission. This threshold may be an overestimate of what will be required, but the other assumptions are optimistic, as it is unlikely that LISA will have full functionality for $5$ years. Results will be based on this LISA configuration (`5yr, 2d') unless explicitly stated otherwise, but this should be regarded as optimistic. Given the uncertainty in data analysis practicalities and on the lifetime of the mission, we will also present results from a pessimistic case in which we assume the mission lasts only two years and there is a partial failure on the satellite which means only one of the low-frequency channels is available (`2yr, 1d').

If EMRI events were short-duration, the event rate would be determined by computing the distance at which the SNR equals $\rho_{\rm thresh}$ and then multiplying the rate per unit volume by the volume contained by that distance. However, EMRIs are long-lived, and SNR can be accumulated for as much of the inspiral as coincides with the LISA observation. For a given system at a given distance it is possible to compute the SNR as a function of the time {\it remaining} until plunge, $\tau_{pl}$. If the LISA mission lasts $T_{\rm LISA}$ years, then for $\tau_{pl} < T_{\rm LISA}$, this SNR is accumulated over $\tau_{pl}$ years, while for $\tau_{pl} > T_{\rm LISA}$, this SNR is accumulated over the $T_{\rm LISA}$ years from $\tau_{pl}=\tau_0$ to $\tau_{pl}=\tau_0 - T_{\rm LISA}$. As a general rule, the SNR will initially increase as $\tau_{pl}$ decreases and then decrease as $\tau_{pl}$ approaches zero. So, there will be a largest (earliest) $\tau_{pl}$ for which the SNR exceeds $\rho_{\rm thresh}$, $\tau_{\rm early}$ say, and a smallest (latest) $\tau_{pl}$, $\tau_{\rm late}$ say. If the LISA satellite turns on when a particular system is at any point in the range $\tau_{\rm late} < \tau_{pl} < \tau_{\rm early}$, that system will be seen by LISA. We can thus define an {\em observable lifetime} for EMRI systems, $\tau$, by
\begin{equation}
\tau ({\bf \lambda}) =  \tau_{\rm early} ({\bf \lambda}) - \tau_{\rm late} ({\bf \lambda})
\end{equation}
where ${\bf \lambda}$ denotes the system parameters. If EMRIs plunge at a rate ${\cal R}$ per year in a particular galaxy, then $\tau {\cal R}$ gives the expected number of events that LISA will observe from that galaxy, after appropriate averaging of ${\cal R}$ over the other system parameters, ${\bf \lambda}$.

\subsection{SNR Calculation}
\label{SNRcalc}
To compute SNRs, we use the fluxes presented in Finn and Thorne~\cite{FT}. They computed the gravitational wave emission from circular and equatorial EMRIs via solution of the Teukolsky equation, i.e., accurate to arbitrary orders in velocity, but only to leading order in mass-ratio. These are the most accurate fluxes presently available for a range of central black hole spins. The restriction to circular and equatorial orbits is unphysical in that most mechanisms predict EMRI orbits will be of moderate eccentricity and inclined to the black hole equatorial plane when they enter the LISA band~\cite{Pau}. However, consideration of circular-equatorial orbits is a necessary first step and should indicate what the dependence of the LISA sensitivity on the central black hole mass and spin will be like, the investigation of which is the primary goal of this paper. Eccentricity and inclination introduce complexity into the waveform which will impact these results. The extension of this work to eccentric and inclined orbits is a necessary future project, which could be undertaken using approximate EMRI waveform models~\cite{AK,NK}.

The SNR, $\rho$, of an EMRI consisting of a compact object of mass $\mu$ falling into a massive black hole of mass $M$ and dimensionless spin $a=S/M^2$ at a redshift $z$ is given by~\cite{FT}
\begin{eqnarray}
\fl\rho^2 &=& \sum_{m=1}^{4} \int \left[ \frac{h_{c,m}^2}{f_m^2 \, S_h^{SA}(f_m)} \right]{\rm d}f_m \nonumber \\
\fl\mbox{where} && h_{c,1} = \frac{5}{\sqrt{672\pi}} \sqrt{\frac{\mu}{M}} \frac{M(1+z)}{d_L(z)} \tilde{\Omega}^{1/6} {\cal H}_{c,1} \nonumber \\
\fl \mbox{and $m\geq2$}&& h_{c,m} = \sqrt{\frac{5(m+1)(m+2)(2m+1)!m^{2m}}{12\pi(m-1)[2^m m!(2m+1)!!]^2}}  \sqrt{\frac{\mu}{M}} \frac{M(1+z)}{d_L(z)} \tilde{\Omega}^{(2m-5)/6} {\cal H}_{c,m}. \label{SNRdef}
\end{eqnarray}
In this, $S_h^{SA}$ denotes the sky and orientation averaged spectral density of the detector, $d_L(z)$ is the luminosity distance to redshift $z$ and $\tilde{\Omega} = M(1+z)\Omega = 1/(a+(r/M)^{3/2})$ is the dimensionless angular frequency of the orbit when the object is at Boyer-Lindquist radius $r$. The ${\cal H}_{c,m}$'s are relativistic correction factors to the post-Newtonian waveform amplitudes that may be computed from tables in~\cite{FT}. The summation over waveform harmonics stops at $m=4$, since no higher harmonics are tabulated in~\cite{FT}. However, the SNR contributed by the $m=4$ harmonic is already a small fraction of the total, and so the inclusion of harmonics with $m\geq5$ will make only a small correction to these results. Finn and Thorne tabulate results only as far as $r/r_{\rm isco} = 10$. Beyond that radius, we use extrapolations based on polynomials in $1/\sqrt{r}$ with the constraint that ${\cal H}_{c,m} \rightarrow 1$ as $r\rightarrow \infty$. We have checked that our results are not sensitive to the exact way in which the extrapolation is done. In this analysis, we used $S_h^{SA}$, including confusion from white-dwarf binaries, taken from~\cite{AK}. For the `5yr, 2d' case, we multiplied the SNR by a factor of $\sqrt{2}$ to account for the fact that LISA can be thought to consist of two independent detectors at low-frequency. We computed $d_L(z)$ using a standard $\Lambda$-CDM cosmology with $H_0 = 71$km s$^{-1}$ Mpc$^{-1}$, $\Omega_M = 0.27$ and $\Omega_{\Lambda}=0.73$. These SNRs use a low-frequency approximation to the LISA response, and so we will slightly underestimate the SNRs for systems with smaller central black hole masses, $M$, since at higher frequencies the LISA response gains a third independent channel.

Expression~\erf{SNRdef} allows us to compute the observable lifetime for EMRIs as a function of $\mu$, $M$, $a$ and $z$. We note that we are computing the observable lifetime of a sky-and-orientation-averaged source, rather than the sky-and-orientation-averaged observable lifetime of a source, which is what we actually need, but we hope these will be comparable.

\subsection{Observable lifetimes}
\label{obslife}
\begin{figure}
\begin{center}
\includegraphics[keepaspectratio=true, width=4in, angle=0]{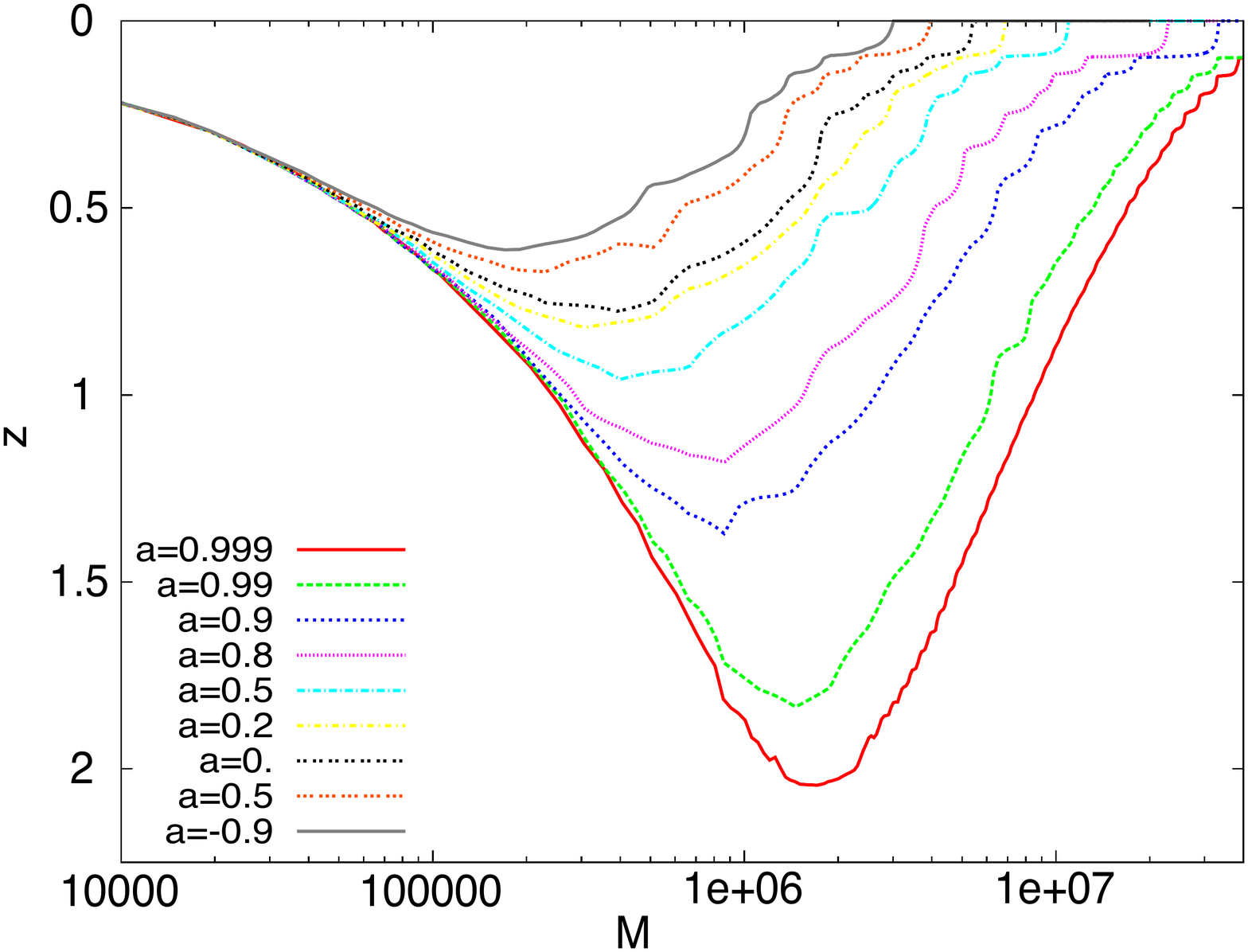}\\
\includegraphics[keepaspectratio=true, width=4in, angle=0]{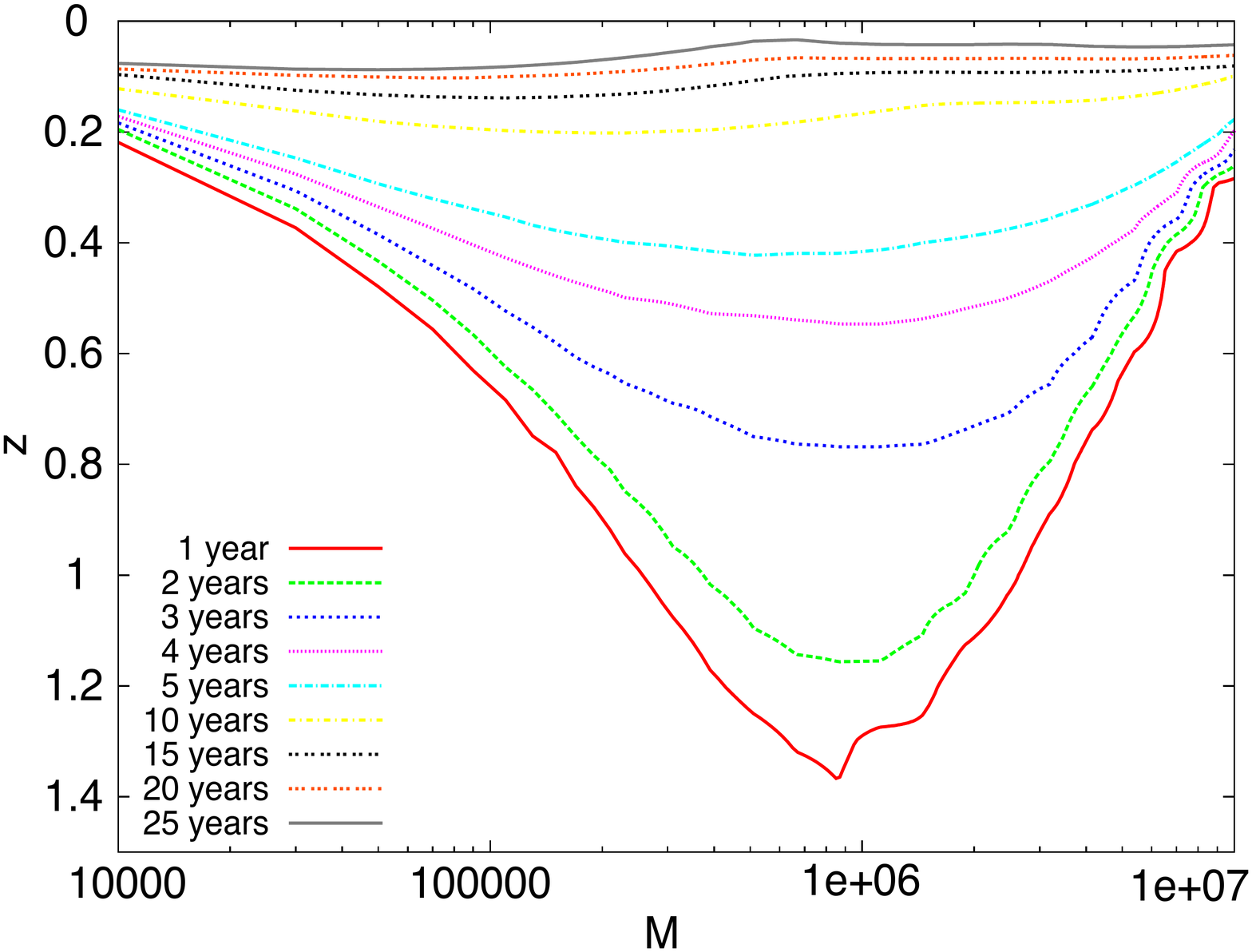}
\end{center}
\caption{Contours of constant observable lifetime for BH inspirals. The upper plot shows contours with $\tau = 1$yr for a range of central black hole spins, while the lower plot shows contours for prograde orbits into central black holes with spin $a=0.9$ for a range of values of $\tau$.}
\label{fig1}
\end{figure}

In Figure~\ref{fig1} we show contours of constant observable lifetime on the $(M,z)$ plane for black hole inspirals with $\mu=10M_{\odot}$. Note that here and elsewhere we quote masses and observable lifetimes as measured at the source as these are of most use when computing event rates, since the rate of EMRIs per black hole, ${\cal R}$, is usually quoted in the source frame. The apparent mass and observable lifetime will be $M(1+z)$ and $\tau (1+z)$. We see that LISA is most sensitive to central black holes with mass around $10^6M_{\odot}$, and can see many systems out to $z\sim1$, with prograde inspirals into rapidly spinning MBHs being visible out to $z\sim 2$. For rapidly spinning black holes, the distance sensitivity is increased and shifted to higher intrinsic MBH masses. This follows from the fact that the emission comes out at higher frequencies for a given black hole mass, and hence the frequency at the floor of the LISA sensitivity curve corresponds to larger $M$.

\begin{figure}
\begin{center}
\includegraphics[keepaspectratio=true, width=4in, angle=0]{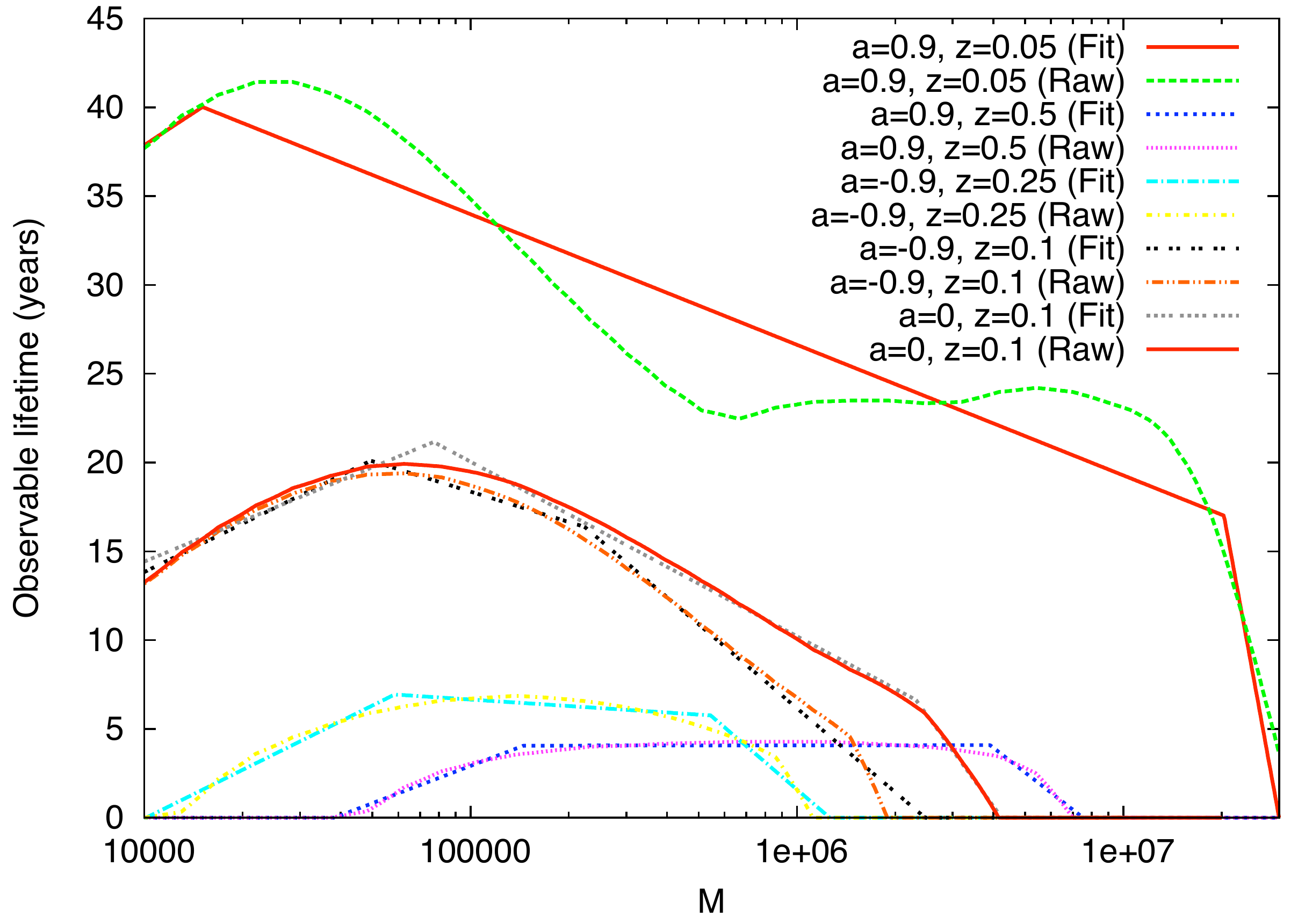}\\
\end{center}
\caption{Trapezium fits to the observable lifetime as a function of central black hole mass, for a variety of spins and source redshifts, as indicated in the legend, and with $\mu=10M_{\odot}$. Lifetimes computed from the numerical observable lifetime data are indicated as ``Raw'' and those computed from the fit by ``Fit''. The values of $a$ and $z$ for this plot were chosen at random, but these curves are representative of all other choices. We see that the simple trapezium fit works well in all cases.}
\label{fig2}
\end{figure}

The lifetime $\tau$, for a fixed redshift $z$, can be well approximated by a trapezium function of $\log(M)$, with break-points at $\log(M) = x_{\rm min}, x_1,x_2,x_{\rm max}$ and lifetimes of $\tau=0,y_1,y_2,0$ at those points; i.e., 
\begin{equation}
\tau = \left\{\begin{array}{ll}0&\mbox{$\log(M) \leq x_{\rm min}$}\\ 
y_1 \, (\log(M)-x_{\rm min})/(x_1-x_{\rm min})& \mbox{$x_{\rm min} \leq \log(M) \leq x_{1}$} \\
y_1+(y_2-y_1) \, (\log(M)-x_{1})/(x_2-x_{1})& \mbox{$x_{1} \leq \log(M) \leq x_{2}$} \\
y_2 \, (1-(\log(M)-x_{2})/(x_{\rm max}-x_{2})) & \mbox{$x_{2} \leq \log(M) \leq x_{\rm max}$} \\
0&\mbox{$x_{\rm max} \leq \log(M)$}\end{array} \right. .
\label{trapfit}
\end{equation}
This is illustrated in Figure~\ref{fig2} for a few cases. For low $z$ and high spin in particular, the observable lifetime shows a lot more features which arise due to the shape of the LISA noise curve. However, we ignore these in the interest of having a simple fitting function. We will see in Figure~\ref{fig3} that the fits reproduce the event distribution well. We have also obtained fits to the six functions $x_{\rm min}, x_1, y_1, x_2, y_2, x_{\rm max}$ as a function of $z$ for the different central black hole spins quoted in~\cite{FT}. The fits are given in Table~\ref{tab1} for $a=0, 0.9, -0.9$. Corresponding fits for the pessimistic detector configuration are given in Table~\ref{tab1b}. Note that here and elsewhere, an inspiral with $a<0$ denotes a {\it retrograde} inspiral into a black hole of spin $|a|$. For other values of the spin, a reasonable approximation may be obtained by linear extrapolation of the lifetimes given here. The functional forms of these fits were chosen arbitrarily, and these functions could probably be fit equally well using a different prescription. Given the other uncertainties about the EMRI population, these fits are probably more accurate than required in practice. However, we have opted to provide an accurate description of the instrumental response (which is less uncertain than the astrophysics) and encourage readers to simplify these fits or not according to the application. We hope that these fits will be useful for computations of LISA EMRI event distributions for particular MBH population models.

The fits in Tables~\ref{tab1}--\ref{tab1b} apply to inspirals of black holes ($\mu=10M_{\odot}$). We provide only these results as such events will dominate the LISA event rate. The observable lifetime is affected by the shape of the LISA noise curve and the imposed SNR threshold, so the scaling with $\mu$ is non-trivial. However, for lower values of $\mu$ the observable lifetime tends to be longer at lower redshift, since the inspiral proceeds more slowly, but the observable lifetime falls off more quickly as the redshift increases, since the instantaneous amplitude is lower and the threshold SNR is reached at a much lower redshift. As an illustration, the observable lifetime (in years) for $a=0.9$ and $M=10^6M_{\odot}$ at $z=\{0.01,0.05,0.1,0.2,0.3,0.4,0.5\}$ is $\{31,23,14,8.7,6.5,5.1,4.3\}$ for $\mu=10M_{\odot}$,  $\{91,43,21,8.3,4,1,0\}$ for $\mu=1.4M_{\odot}$ and $\{187,40,14,0.6,0,0,0\}$ for $\mu=0.6M_{\odot}$. For $M=3\times10^5$ these numbers become $\{33,26,17,10,6.8,5,4\}$ for $\mu=10M_{\odot}$,  $\{111,42,16,4,0,0,0\}$ for $\mu=1.4M_{\odot}$ and $\{197,28,4.5,0,0,0,0\}$ for $\mu=0.6M_{\odot}$.

\begin{table}
\begin{tabular}{|c|c|c|}
\hline Spin&Coefficient&Fit \\\hline
&$x_{\rm min}$&$-16.1+602\,z-602\,z^2/(0.05+z)$ \\\cline{2-3}
&$x_{1}$&$9.5+40.0\,z-38.6\,z^2/(0.05+z)$ \\\cline{2-3}
$0$&$y_{1}$&$3.3-0.64/(1-z)+(0.28+0.22\,z)/(0.005+z^2)$ \\\cline{2-3}
&$x_{2}$&$16.3-49.2\,z+47.8\,z^2/(0.05+z)$ \\\cline{2-3}
&$y_{2}$&$8.4-9.7\,z$ \\\cline{2-3}
&$x_{\rm max}$&$15.8-10.2\,z+7.7\,z^2/(0.05+z)$ \\\hline
&$x_{\rm min}$&$-7.7+395\,z-394\,z^2/(0.05+z)$ \\\cline{2-3}
&$x_{1}$&$10.8+2\,z+0.09/(1+z)$ \\\cline{2-3}
$0.9$&$y_{1}$&$-0.15+5.9\,\exp(-z)-0.03/(1.45-z)+(0.26-0.06\,z)/(0.005+z^2)$ \\\cline{2-3}
&$x_{2}$&$13.5-0.62\,z+3/(1+z)$ \\\cline{2-3}
&$y_{2}$&$0.78+5.1\,\exp(-z)-0.15/(1.45-z)+(0.08+0.01\,z)/(0.005+z^2)$ \\\cline{2-3}
&$x_{\rm max}$&$14.5-0.75\,z+2.6/(1+z)$ \\\hline
&$x_{\rm min}$&$60.8-22.3\,z-57.9/(1+z)$ \\\cline{2-3}
&$x_{1}$&$7.1+4.1\,z+3.6/(1+z)$ \\\cline{2-3}
$-0.9$&$y_{1}$&$-1.2+5.9\,\exp(-z)-0.12/(0.7-z)+(0.28-0.08\,z)/(0.005+z^2)$ \\\cline{2-3}
&$x_{2}$&$15.4-43.5\,z+41.8\,z^2/(0.05+z)$ \\\cline{2-3}
&$y_{2}$&$8.1-11\,z$ \\\cline{2-3}
&$x_{\rm max}$&$15.1-3.5\,z-0.07/(1+z)$ \\\hline
\end{tabular}
\caption{Fitting functions for coefficients in the trapezium fit to the observable lifetime described by Eq.~\erf{trapfit}. The fits are valid for $z <1$, $z<1.45$ and $z<0.7$ for $a=0,0.9,-0.9$ respectively. For $z$ outside this range, $\tau=0$.}
\label{tab1}
\end{table}

\begin{table}
\begin{tabular}{|c|c|c|}
\hline Spin&Coefficient&Fit \\\hline
&$x_{\rm min}$&$4.0+124\,z-119\,z^2/(0.05+z)$ \\\cline{2-3}
&$x_{1}$&$10.3+ 24.4 z - 21.8 \,z^2/(0.05+z)$ \\\cline{2-3}
$0$&$y_{1}$&$0.5-0.03/(0.55- z) + (0.13- 0.041\,z)/(0.005+z^2)$ \\\cline{2-3}
&$x_{2}$&$10.7+ 74.5\,z - 77.2\,z^2/(0.05+z)$ \\\cline{2-3}
&$y_{2}$&$-2.5 + 5.3\,\exp(-z) - 0.02/(0.55- z) + (0.12-0.23\,z)/(0.005+ z^2)$ \\\cline{2-3}
&$x_{\rm max}$&$15.7- 7.85\,z + 3.92\,z^2/(0.05+z)$ \\\hline
&$x_{\rm min}$&$3.76+ 132\,z - 127.9\,z^2/(0.05+z)$ \\\cline{2-3}
&$x_{1}$&$19.9- 1.84\,z - 8.81/(1 + z)$ \\\cline{2-3}
$0.9$&$y_{1}$&$0.45+ 0.45\,\exp(-z)- 0.08/(1- z) + (0.12+0.21\,z)/(0.005+z^2)$ \\\cline{2-3}
&$x_{2}$&$9.51+ 1.23\,z + 6.77/(1 + z)$ \\\cline{2-3}
&$y_{2}$&$-0.8 + 3.1\,\exp(-z) - 0.06/(1- z) + (0.04+0.28\,z)/(0.005+z^2)$ \\\cline{2-3}
&$x_{\rm max}$&$15.1- 1.77\,z + 1.84/(1 + z)$ \\\hline
&$x_{\rm min}$&$82.5- 39.7\,z - 77.3/(1 + z)$ \\\cline{2-3}
&$x_{1}$&$17.6- 1.1\,z - 7.1/(1 + z)$ \\\cline{2-3}
$-0.9$&$y_{1}$&$1.9- 2.5\,\exp(-z) - 0.01/(0.42- z)+ (0.14- 0.04\,z)/(0.005+z^2)$ \\\cline{2-3}
&$x_{2}$&$11.7+ 10.9\,z - 24.3 z^2$ \\\cline{2-3}
&$y_{2}$&$1.1- 1.4\,\exp(-z) - 0.009/(0.42-z) + (0.12 + 0.006\,z)/(0.005+ z^2)$ \\\cline{2-3}
&$x_{\rm max}$&$18.0- 7.0\,z - 3.2/(1 + z)$ \\\hline
\end{tabular}
\caption{As Table~\ref{tab1}, but now for the pessimistic detector case, `2yr, 1d'. The fits are valid for $z <0.55$, $z<1$ and $z<0.42$ for $a=0,0.9,-0.9$ respectively. For $z$ outside this range, $\tau=0$.}
\label{tab1b}
\end{table}

\section{LISA EMRI event rates}
\label{LISAevents}
To compute the number and parameter distribution of events that LISA will see, we need a prescription for the comoving number density of MBHs, ${\cal N}(M,a,z)$, and the rate of EMR inspirals in any particular system, ${\cal R}(M,a,z)$. The quantity ${\cal N}(M,a,z) \rmd M\rmd a$ denotes the number of massive black holes per comoving volume with mass in the range $M\rightarrow M+\rmd M$ and spin in the range $a\rightarrow a+\rmd a$. As little is known about black hole spins, we assume that the MBH mass and spin distributions are independent and factorize ${\cal N}(M,a,z) \rmd M\rmd a = (\rmd n/\rmd\ln M)(M,z)\,\rmd\ln M\,p(a,z)\rmd a$, with normalisation $\int p(a,z)\rmd a=1$. For simplicity we further assume that there is no evolution in the properties of black holes with $z$, so that $ (\rmd n/\rmd\ln M)(M,z) =  (\rmd n/\rmd\ln M)(M)$ and $p(a,z)=p(a)$ only. This is probably reasonable for the range of redshifts of LISA EMRI events. The number of LISA events is then given by
\begin{equation}
\fl N_{\rm LISA} = \int_{z=0}^\infty \int_{M=M_{\rm low}}^{M_{\rm high}} \int_{a=-1}^{1} {\cal R}(M) \, \tau(M,a,z) \,\frac{\rmd n}{\rmd \ln M} (M)\,p(a)\,\frac{\rmd V_c}{\rmd z}\,\rmd a\,\rmd \ln M \, \rmd z.
\label{nevent}
\end{equation}
Here $(\rmd V_c/\rmd z) \,\rmd z$ is the comoving volume in the redshift range $z$ to $z+\rmd z$, which we compute for the same $\Lambda$-CDM cosmology used previously. 

The mass function, $\rmd n/\rmd\ln(M)$, can be derived from observed galaxy luminosity functions using the $L-\sigma$ and $M-\sigma$ relations. Using results from Aller \& Richstone~\cite{AR}, one finds the function $\rmd n/\rmd \ln M$ is approximately flat and equal to $3.6\times 10^{-3} h_{71}^2$Mpc$^{-3}$ for MBHs in the LISA mass range, $M < 5\times10^6M_{\odot}$ (NB $h_{71} = H_0/(71$km s$^{-1}$ Mpc$^{-1}$)). If we remove Sc-Sd galaxies (several of these are known to have black holes of much lower mass than that derived from the luminosity~\cite{desroches08}), this becomes approximately  $1.8\times 10^{-3} h_{71}^2$Mpc$^{-3}$. This was the expression used to derive LISA EMRI event rates in~\cite{gairetal}. However, there is great uncertainty in the black hole mass function in the LISA range, since there are very few measurements of the masses of quiescent black holes in that range~\cite{shankar07,sesana}. We therefore adopt the prescription
\begin{equation}
\frac{\rmd n}{\rmd \ln M} = n_{0} \left(\frac{M}{3\times 10^6 M_{\odot}}\right)^\alpha
\end{equation}
and will normalise to the case $n_{0}=0.002$Mpc$^{-3}$ and $\alpha=0$.

We can adopt a similar ansatz for the intrinsic EMRI rate per galaxy, i.e., the frequency with which EMR inspirals start or end within a particular system
\begin{equation}
{\cal R}_{\gamma} = {\cal R}^{\gamma}_{\rm MW} \left(\frac{M}{3\times 10^6 M_{\odot}}\right)^{\beta_\gamma}.
\end{equation}
Here $\gamma$ denotes the type of EMRI  --- BH, NS or WD  (which we take to have masses $\mu=10M_{\odot}, 1.4M_{\odot}, 0.6M_{\odot}$ respectively) --- and the subscript `MW' indicates we are normalising to a mass comparable to the Milky Way black hole. Hopman~\cite{hopmanLISA} quotes rates and scalings of ${\cal R}^{\gamma}_{\rm MW} = {\cal R}^{\gamma}_0 = 400$Gyr$^{-1}, 7$Gyr$^{-1}, 20$Gyr$^{-1}$ and $\beta_{\gamma} = -0.15,-0.25,-0.25$ for inspiraling black holes, neutron stars and white dwarfs respectively. The subscript `$0$' is just used to distinguish these fixed reference rates from our chosen ${\cal R}^{\gamma}_{\rm MW}$. There is presently no observational prescription for the MBH spin distribution, $p(a)$, so we use the simple ansatz $p = (\delta(a-a_0) + \delta(a+a_0))/2$ for three cases $a_0=0,0.5,0.9$. This assumes prograde and retrograde inspirals are equally likely in a system with given spin.

Putting this together we can quote the number of events per species as
\begin{equation}
N_{\gamma} = N^{\gamma}_0 \left(\frac{{\cal R}^{\gamma}_{\rm MW}}{{\cal R}^{\gamma}_{0}} \right) \,f_{\gamma} (\alpha+\beta_\gamma)\, \left( \frac{n_{0}}{0.002{\rm Mpc}^{-3}}\right) .
\label{events}
\end{equation}
The coefficients $N^{\gamma}_0$ and functions $f_\gamma$ are tabulated in Table~\ref{tab2} for both the optimistic and pessimistic detector configurations. We normalise the rates per galactic black hole to the reference rates of Hopman~\cite{hopmanLISA} and normalise the $f_\gamma$'s such that $f_\gamma = 1$ when $\alpha+\beta_\gamma = -0.15, -0.25, -0.25$ for BHs, NSs and WDs respectively. Note that in this simple prescription we cannot disentangle the effects of $\alpha$ and $\beta_\gamma$ on the event rate. We note also that we only include MBHs in the range $10^4 M_{\odot} < M < 3\times 10^7 M_{\odot}$ in our event rate computation
. The main conclusion from this table is that the event rate is very high and is totally dominated by BH inspirals, since these are visible much further than NS or WD inspirals, and are intrinsically more common in the reference model. Only if the WD or NS rates are enhanced by other mechanisms (e.g., tidal stripping of binaries~\cite{Pau}) or rise much more steeply toward low $M$ will we see significant numbers of those events. Using the pessimistic detector assumptions, the rate decreases by approximately a factor of $10$, although the rate scaling, $f_\gamma$, does not change significantly. The event rates in the optimistic detector case are quite large compared to previous EMRI event rate estimates~\cite{gairetal}. This is primarily because earlier work took the rate scaling to be $\beta_\gamma = 3/8$ for all species, which suppresses the number of events with $M < 3\times 10^6M_{\odot}$ and these are the dominant contributors to our event rate. With $\beta_\gamma = 3/8$ and $a=0.9$ we predict $0.7, 1.2$ and $650$ events for WDs, NSs and BHs respectively. The number of WD events ($\sim100$) predicted in~\cite{gairetal} was two orders of magnitude greater than the result we quote here. This arose because this previous work used a much higher WD rate of $5$Myr$^{-1}$, which was based on older simulations. All of these numbers should be treated with caution, as there are very large uncertainties in the intrinsic rate of inspirals per galaxy, and the rates we have used could be off by as much as two orders of magnitude~\cite{Pau}. At a rate of $4$ BH inspirals per Gyr per galaxy, we will see only $10$--$20$ events over the mission lifetime, even with optimistic assumptions about the detector.

\begin{table}
\begin{tabular}{|c|c|c|c|}
\hline
Spin&Species&$N^{\gamma}_0$&$f_\gamma(x=\alpha+\beta_\gamma)$ \\\hline
&BH&$\begin{array}{c}1000\\\left[150\right]\end{array}$&$\begin{array}{c}\exp(-0.42-2.70x + 0.57 x^2)\\\left[\exp(-0.40 - 2.60 x + 0.53 x^2)\right]\end{array}$ \\\cline{2-4}
0&NS&$\begin{array}{c}1.2\\\left[0.1\right]\end{array}$&$\begin{array}{c}\exp(-0.64 -2.43 x + 0.49 x^2)\\\left[\exp(-0.58 - 2.24 x + 0.39 x^2)\right]\end{array}$ \\\cline{2-4}
&WD&$\begin{array}{c}0.6\\\left[0.06\right]\end{array}$&$\begin{array}{c}\exp(-0.62 -2.36 x + 0.47 x^2)\\\left[\exp(-0.57 - 2.18 x + 0.36 x^2)\right]\end{array}$ \\\hline
&BH&$\begin{array}{c}1100\\\left[170\right]\end{array}$&$\begin{array}{c}\exp(-0.39 -2.54 x + 0.64 x^2)\\\left[\exp(-0.37 - 2.39 x + 0.60 x^2)\right]\end{array}$ \\\cline{2-4}
0.5&NS&$\begin{array}{c}1.4\\\left[0.1\right]\end{array}$&$\begin{array}{c}\exp(-0.56 -2.08 x + 0.57 x^2 - 0.081 x^3)\\\left[\exp(-0.51 - 1.93 x + 0.46 x^2 - 0.044 x^3)\right]\end{array}$ \\\cline{2-4}
&WD&$\begin{array}{c}0.75\\\left[0.07\right]\end{array}$&$\begin{array}{c}\exp(-0.53-1.98 x + 0.54 x^2 - 0.081 x^3)\\\left[\exp(-0.49 - 1.86 x + 0.43 x^2 - 0.037 x^3)\right]\end{array}$ \\\hline
&BH&$\begin{array}{c}1580\\\left[260\right]\end{array}$&$\begin{array}{c}\exp(-0.30-1.85 x + 0.82 x^2 - 0.073 x^3)\\\left[\exp(-0.26 - 1.60 x + 0.77 x^2 - 0.10 x^3)\right]\end{array}$ \\\cline{2-4}
0.9&NS&$\begin{array}{c}2.6\\\left[0.3\right]\end{array}$&$\begin{array}{c}\exp(-0.36-1.27 x + 0.70 x^2 - 0.14 x^3)\\\left[\exp(-0.32 - 1.11 x + 0.59 x^2 - 0.10 x^3)\right]\end{array}$ \\\cline{2-4}
&WD&$\begin{array}{c}1.4\\\left[0.1\right]\end{array}$&$\begin{array}{c}\exp(-0.33-1.13 x + 0.66 x^2 - 0.14 x^3)\\\left[\exp(-0.29 - 1.02 x + 0.54 x^2 - 0.091 x^3)\right]\end{array}$ \\\hline
\end{tabular}
\caption{Normalisation factors and functions appearing in Eq.~\erf{events}. The dependence on the rate scaling, $\alpha+\beta_\gamma$, was obtained by empirical fitting with simple functions in the range $-1.5 \leq x \leq 1.5$ and may not be trustworthy outside this range. The second line in each row, denoted by square brackets, is the corresponding result for the pessimistic `2yr, 1d' configuration.
}
\label{tab2}
\end{table}

\begin{figure}
\begin{center}
\includegraphics[keepaspectratio=true, width=3.8in, angle=0]{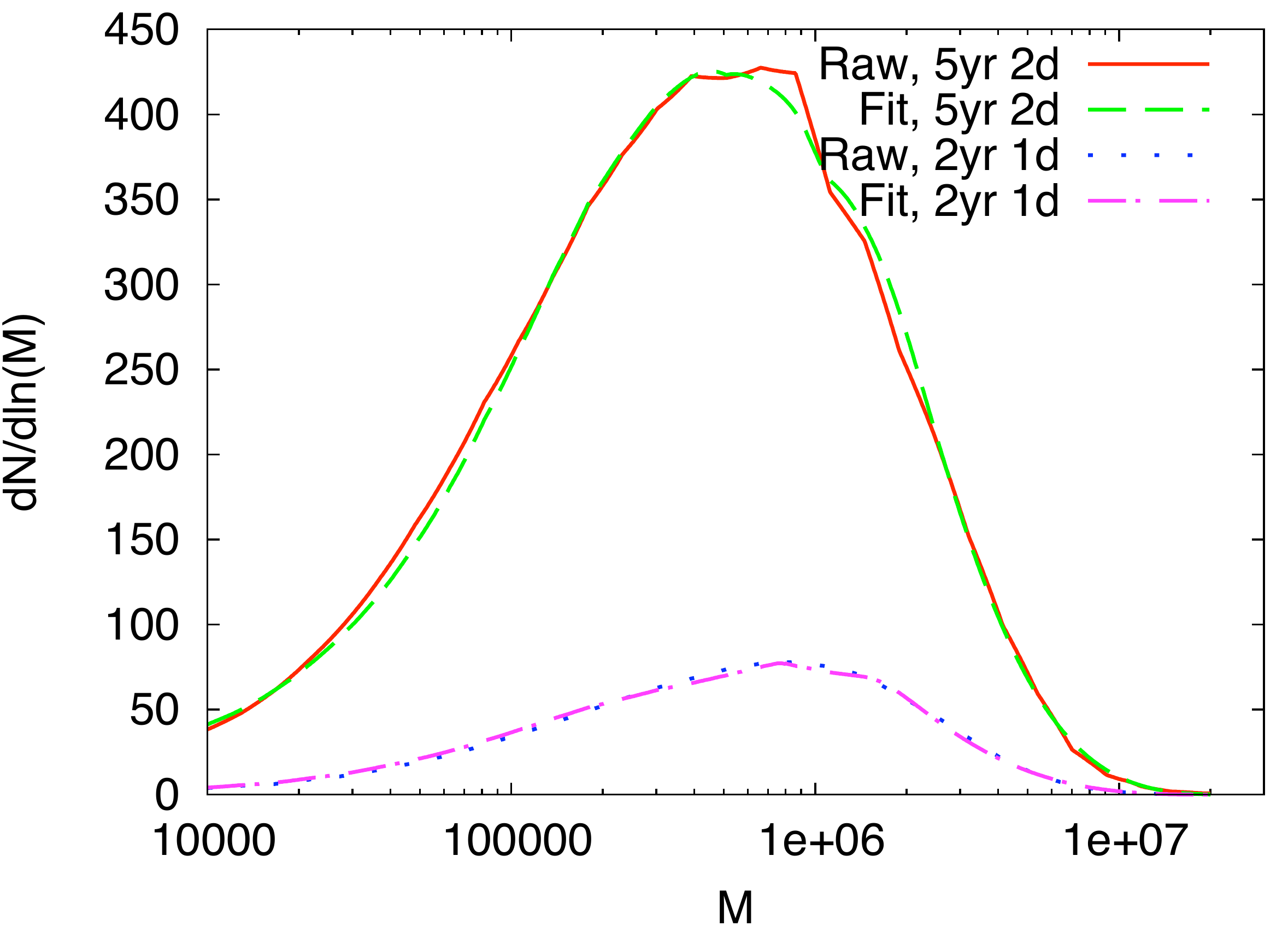}\\
\includegraphics[keepaspectratio=true, width=3.8in, angle=0]{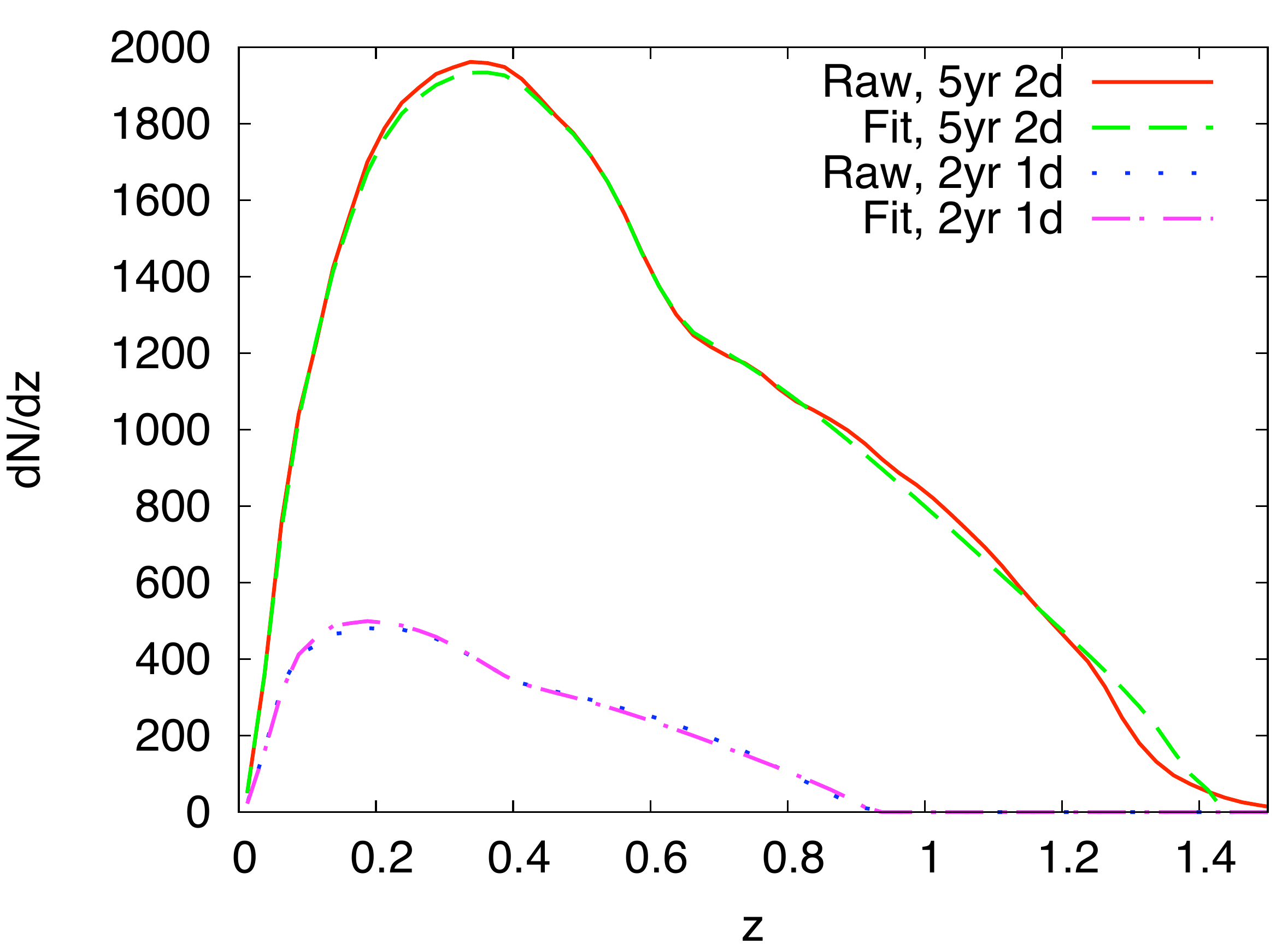}
\end{center}
\caption{Distribution of LISA EMRI events as a function of the intrinsic central black hole mass (upper plot) and as a function of redshift (lower plot), for an inspiraling black hole with $\mu=10M_{\odot}$. We show distributions for both detector configurations, computed using both the raw data and the fitting functions given earlier, as labelled.}
\label{fig3}
\end{figure}

In Figure~\ref{fig3} we show the distribution of events as a function of $M$ and $z$ for the BH events in our reference system with $a_0=0.9$. The distribution is peaked at a mass slightly below $10^6M_{\odot}$, and events are mostly confined to the range $10^5M_{\odot} < M < 2\times 10^6 M_{\odot}$, although there is a significant tail toward smaller central black holes. The redshift distribution is peaked at $z\sim0.4$ with a long tail out to redshifts $z>1$. With pessimistic detector assumptions, the distribution is shifted to lower intrinsic masses and lower redshifts, and with a factor of ten fewer events overall. A useful quantity to compute is the value of $z$, $z_{\rm min}$, such that the expected number of events with $z < z_{\rm min}$ is equal to $1$. This is an estimate of the lowest redshift source that we might observe. The highest likely redshift of a source can be defined in a similar way. We tabulate these quantities for several different BH EMRI rates in Table~\ref{tab3}. It is clear that we expect most of our events to be in the range $0.1 < z < 1$, but if MBHs tend to have high spins and the intrinsic EMRI rate per MBH is also high, we could see an event with redshift as high as $z=1.5$ or as low as $z=0.02$. We note that the minimal likely redshift is independent of the central black hole spin. This is because nearby sources are visible for longer, and inspirals spend proportionally longer far from the central black hole. The nearest source detected is therefore likely to be observed early, i.e., long before plunge, in which phase the spin of the black hole plays a minor role. With pessimistic detector assumptions, the amount of observable inspiral is decreased, so the minimal redshift is larger, but still largely independent of assumptions on the black hole spin.

\begin{table}
{\footnotesize\begin{tabular}{|c|c|c|c|c|c|c|c|c|c|}
\hline
Spin&\multicolumn{3}{c|}{0}&\multicolumn{3}{c|}{0.5}&\multicolumn{3}{c|}{0.9}\\\hline
${\cal R}^{\rm BH}_{\rm MW} $(Gyr$^{-1}$)&4&40&400&4&40&400&4&40&400\\\hline
$z_{min}$ \begin{tabular}{c}5yr, 2d\\ 2yr, 1d\end{tabular}&$\begin{array}{c}0.14\\0.30^{*}\end{array}$&$\begin{array}{c}0.05\\ 0.07\end{array}$&$\begin{array}{c}0.02\\ 0.03\end{array}$&$\begin{array}{c}0.14\\0.29^{*}\end{array}$&$\begin{array}{c}0.05\\0.07\end{array}$&$\begin{array}{c}0.02\\0.03\end{array}$&$\begin{array}{c}0.14\\0.27^{*}\end{array}$&$\begin{array}{c}0.05\\0.07\end{array}$&$\begin{array}{c}0.02\\0.03\end{array}$\\\hline
$z_{max}$\begin{tabular}{c}5yr, 2d\\ 2yr, 1d\end{tabular}&$\begin{array}{c}0.64\\0.16^{*}\end{array}$&$\begin{array}{c}0.76\\0.41\end{array}$&$\begin{array}{c}0.84\\0.49\end{array}$&$\begin{array}{c}0.74\\0.19^{*}\end{array}$&$\begin{array}{c}0.94\\0.49\end{array}$&$\begin{array}{c}1.0\\0.59\end{array}$&$\begin{array}{c}1.1\\0.41^{*}\end{array}$&$\begin{array}{c}1.3\\0.76\end{array}$&$\begin{array}{c}1.4\\0.86\end{array}$\\\hline
\end{tabular}}
\caption{Most likely minimum and maximum redshift LISA events that will be detected, as a function of the central black hole spin assumed, and the assumed rate of EMR inspirals in $M=3\times10^6M_{\odot}$ black holes. We show results for both detector configurations. The cases indicated by a $*$ have $z_{min} > z_{max}$, but this is an artefact because in those cases less than 2 events are observed in total.}
\label{tab3}
\end{table}
It has been suggested that EMRI observations could be used to make high precision measurements of the Hubble constant in a statistical way~\cite{mcleod}. However, these measurements would require LISA to detect $\sim10$ events at $z \lesssim 0.23$ or $\sim 20$ events at $z \lesssim 0.5$. For the EMRI population we are considering here, assuming all MBHs have spin $a=0.9$, we find that a rate ${\cal R}^{\rm BH}_{\rm MW} > 14$Gyr$^{-1}$ will produce $>10$ events at $z \lesssim 0.23$, and ${\cal R}^{\rm BH}_{\rm MW} > 10$Gyr$^{-1}$ will produce $>20$ events at $z \lesssim 0.5$. If we assume all MBHs are non-spinning, $a=0$, these increase to $15$Gyr$^{-1}$ and $11$Gyr$^{-1}$ respectively. The corresponding rates for the pessimistic detector configuration are $43$Gyr$^{-1}/41$Gyr$^{-1}$ for $a=0.9$ and $48$Gyr$^{-1}/56$Gyr$^{-1}$ for $a=0$. Taking $\beta_{\rm BH}=0$ rather than $\beta_{\rm BH}=-0.15$ these numbers become $21$Gyr$^{-1}/14$Gyr$^{-1}$ ($a=0.9$) and $24 $Gyr$^{-1}/16$Gyr$^{-1}$ ($a=0$) or  $61$Gyr$^{-1}/54$Gyr$^{-1}$ ($a=0.9$) and $73$Gyr$^{-1}/83$Gyr$^{-1}$ ($a=0$) for the pessimistic detector configuration. These rates are all smaller than the current best estimates in~\cite{hopmanLISA}, which indicates that this measurement should be feasible, although we start to face problems if the intrinsic rate is a factor of $5$ lower and we have the pessimistic detector. 

It has also been suggested that WD EMRIs could be used to make Hubble constant measurements, if an electromagnetic counterpart is observed from the tidal disruption of the WD~\cite{sesana}. This requires us to observe a WD inspiraling into a black hole with $M\lesssim 10^5 M_{\odot}$. Our prescription predicts only $0.1$ WD events in the range $10^4 M_{\odot} \lesssim M\lesssim 10^5 M_{\odot}$, which means it is unlikely that we would see an event. 
However, given the huge astrophysical uncertainties, we should not rule these out completely.

\section{Using LISA as an astrophysical probe}
\label{app}
For the simple model described in the previous section, we have seen that LISA should detect many EMRI events out to redshift $z \sim 1.5$. Each LISA event should provide us with the system parameters to high precision~\cite{AK}, and hence we will determine the masses and spins of as many as $1000$ MBHs in the nearby Universe. Even if the intrinsic inspiral rate per black hole is much less or the LISA detector performs sub-optimally, we should characterize a few tens of low redshift MBHs. This information can be used for astrophysics. Due to the uncertainties in the rate of EMRIs per galaxy, the total number of events is not a good probe of the MBH population, but the distribution of events as a function of mass and/or spin is a useful probe, modulo the $\alpha+\beta_\gamma$ degeneracy discussed earlier. It is for this application that the research described here was started and where the interesting astrophysics lies. In order to quantify what LISA can do, we are using the observable lifetime functions described here, in conjunction with results from black hole merger trees~\cite{volon08} to investigate the dependence of the LISA event distribution on the MBH population model. Crudely, if we have as many as $1000$ events, then we can imagine dividing the mass range into ten bins, such that we would expect $100$ events in each bin with our baseline model. The Poisson error in each bin would be $\sim10$, and so if the Universe differed from our model in a way that changed the number of events in a bin by more than $10$ we might hope to detect that in our observations. 
We are currently comparing different galaxy evolution models in a Bayesian way, by evaluating the likelihood that a given observed population of EMRI events came from a Universe with a particular history. This will allow an estimation of the number of EMRI events that will need to be observed in order to make a strong statistical statement about which model best describes the real Universe. This work is in progress, but we hope that the observable lifetime fits that we have presented in this paper will allow others to follow a similar research programme without the overhead of doing the LISA response calculations.

Our present results have some limitations and we plan to quantify these in the future. 
One important question is how many of these events are ``useful'' in the sense that they give good parameter estimation. Naively we would hope that any event detected would have reasonable parameter estimation for the intrinsic parameters, but some nearby systems might be observed several years before plunge and so might not show enough evolution over our observation to allow accurate parameter determination. The fraction of the events that are useful should be quantified by a Fisher Matrix analysis. Another issue is that our waveform model is restricted since we consider only circular-equatorial orbits. It is necessary to check how representative these are by considering observable lifetimes for eccentric and inclined orbits. As perturbative waveforms for generic orbits are computationally very expensive, a first cut at this problem would involve computing these lifetimes using approximate waveforms~\cite{AK,NK}. In addition, in the present work we are computing and using the observable lifetime of a sky-averaged source, but it should be verified explicitly that this does not differ too greatly from the sky-averaged observable lifetime of a source, which is what is really required. Finally, these SNRs are based on the low-frequency approximation to the LISA response and hence will not be completely accurate for systems with particularly low mass or high spin central black holes. This error needs to be quantified by computing more accurate SNRs for those systems using the full LISA response, although we do expect our current results to be conservative in this regard.

\ack I thank Marta Volonteri for useful discussions and ongoing collaboration on this project. This work was supported by a Royal Society University Research Fellowship.

\end{document}